\documentclass[aps,prl,twocolumn,showpacs,showkeys,groupedaddress]{revtex4}
\usepackage{graphicx}
\begin{document}

%\preprint{CAS-ITP-2006-212-03}

\title{Collapse Transition of Two-Dimensional Flexible and Semiflexible Polymers
\footnote{Citation information: H. Zhou, J. Zhou, Z.-C. Ou-Yang, and S. Kumar,
Phys. Rev. Lett.~{\bf 97}, 158302 (2006).}}

\author{Haijun Zhou$^{1}$, Jie Zhou$^1$, Zhong-Can Ou-Yang$^1$, and Sanjay Kumar$^{2}$}

\affiliation{$^1$Institute of Theoretical Physics, the Chinese Academy of Sciences, 
  Beijing 100080, China}
\affiliation{$^2$Department of Physics, Banaras Hindu University, 
  Varanasi 221 005, India}

\date{\today}

\begin{abstract}
  The nature of the globule-coil transition of surface-confined polymers
  has been an issue of debate. Here this 2D collapse transition
  is studied through a partially directed lattice model. 
  In the general case of polymers with positive bending stiffness ($\Delta>0$), 
  the collapse transition is {\em first-order}; it becomes 
  {\em second-order} only in the limiting case of zero bending stiffness 
  ($\Delta\equiv 0$).   These analytical results are confirmed by 
  Monte  Carlo simulations. We also suggest some possible future  experiments.
\end{abstract}


\maketitle

%%% Introduction %%%

The collapse (globule-coil) transition of a self-attracting
chain is one of the fundamental problems in polymer physics
\cite{desCloizeaux-Jannink-1990,Grosberg-Khokhlov-1994,Vanderzande-1998}.  
Being deeply connected with biophysical problems such as DNA condensation, 
chromatin organization, and protein folding, the collapse transition has
also biological relevance. Generally speaking, this transition is caused by the 
competition between monomer-monomer attraction and configurational entropy:
Formation of contacts lowers the energy, but it
requires monomers to be aligned and close to each other, thereby decreasing 
the polymer's degrees of configurational freedom.  When the monomer-monomer attraction
is much stronger than thermal energy, the polymer takes globular 
conformations to maximize the number of contacts.  At the other limit of high 
temperatures, 
the polymer is in completely disordered coil states with maximal entropy. 
At zero external force, the collapse transition occurs at 
a so-called $\theta$-temperature \cite{desCloizeaux-Jannink-1990}.

The exact nature of the collapse transition, however, is not yet completely settled,
despite decades of extensive efforts 
\cite{desCloizeaux-Jannink-1990,Grosberg-Khokhlov-1994,Vanderzande-1998}. 
This transition can be studied by mapping to the tricritical point of the
$\phi^4$-$\phi^6$ $O(n)$ field theory at the ${n}{\to}{0}$ limit 
\cite{deGennes-1975,Stephen-1975,Duplantier-1982}, and is expected to be
second-order in 2D and beyond (see, e.g., \cite{Owczarek-Prellberg-2000});
furthermore, exponents of the temperature-driven 
collapse and adsorption of a 2D polymer grafted on a linear boundary was
obtained  by analytical calculations \cite{Vanderzande-etal-1991,Duplantier-2003}. 
However, experimentally observed collapse transitions of both relatively
flexible \cite{Wang-etal-1998} and semiflexible \cite{Yoshikawa-etal-1996} 
3D polymers are more like first-order transitions. Recent Monte Carlo (MC)
simulation \cite{Rampf-etal-2005} is also in favor of the latter interpretation.

With the advancement in single-molecule force manipulation methods, now
the globule-coil transition can also be induced by external stretching 
\cite{ClausenSchaumann-etal-2000,Baumann-etal-2000}.
According to mean-field theory \cite{Halperin-Zhulina-1991} the 
force-induced collapse transition
is first-order in all dimensions. This claim is confirmed by MC simulation
in 3D \cite{Grassberger-Hsu-2002}, while in 2D the transition is argued to
be second-order by MC simulation \cite{Grassberger-Hsu-2002}
and scaling analysis \cite{Marenduzzo-etal-2003,Rosa-etal-2003}. 

\begin{figure}[b]
  \includegraphics[width=0.6\linewidth]{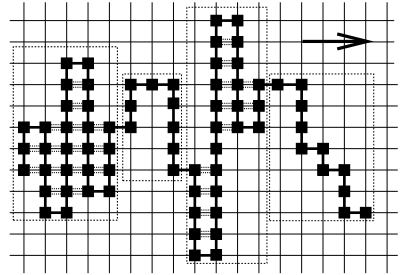}
  \caption{\label{fig:model} 
    The 2D lattice polymer model with contacting interaction, bending stiffness, 
    and external stretching. The arrow shows  the ${\bf z}_0$ direction.
  }
\end{figure}

In order to understand the collapse transition more deeply, here we
investigate a 2D partially directed lattice model of a polymer chain
 which is exactly solvable
(Fig.~\ref{fig:model}). 
As the free energy density of this model system 
can be calculated precisely, we are able to
draw definite conclusions concerning the nature of the collapse transition. 
We find that the collapse transition is second-order for a 
polymer with exactly zero bending stiffness, while it changes 
to be first-order when the bending
stiffness $\Delta$ is non-zero. Therefore bending energies have a dramatic 
effect on the cooperativity of the 2D globule-coil transition. 
We also substantiate our predictions by performing extensive MC simulations.
It becomes experimentally feasible to confine polymers to 2D mobile surfaces 
(see, e.g., \cite{Maier-Raedler-1999}), therefore, future experiments will 
be able to verify  the theoretical predictions. The present work is also
directly applicable to the  biology-related  problem of protein
$\beta$-sheet unfolding.

%%% The model %%%
{\em The model}.---The 2D partially directed polymer 
of $N$ identical units on a square lattice is shown in Fig.~\ref{fig:model}. 
The length of the bond connecting two consecutive monomers $i$ and ${i}{+}{1}$
is fixed to $a_0$, while the direction of bond in ${-}{\bf z}_0$ is prohibited.
If any two monomers $i$ and ${i}{+}{m}$ (${m}{\geq}{3}$) occupy nearest 
neighboring lattice sites, an attractive energy of magnitude $\epsilon$ is gained.
Usually real polymers are semiflexible, we associate an energy penalty of 
magnitude $\Delta$ to each local direction change of the chain 
\cite{Doniach-etal-1996,Kumar-Giri-2005}. 
For this model, only the special case of zero bending stiffness 
has been studied analytically \cite{Brak-etal-1992,Owczarek-Prellberg-1994,Rosa-etal-2003}. 
The results of the present paper suggest that, 
in the general case of ${\Delta}{>}{0}$ the properties of the
polymer will be dramatically different from these earlier predictions
\cite{Brak-etal-1992,Owczarek-Prellberg-1994,Rosa-etal-2003}.

{\em Partition functions}.---To calculate the free energy density of the polymer,
a given configuration of the 2D chain is divided into
a linear sequence of $\beta$-sheet segments and coil segments \cite{Lifson-1964}.
A  $\beta$-sheet segment is defined as a folded segment of ${n_{\beta}}{\geq}{2}$ consecutive columns, 
in which contacting interactions exist between any two adjacent ones. 
Two consecutive $\beta$-sheet segments are separated by a coil segment,
which is a segment of ${n_{\rm c}}{\geq}{0}$ consecutive columns in which all 
monomers are free of
contacts.  For example, the configuration shown in Fig.~\ref{fig:model}
has two $\beta$-sheets and two coils.  
After having made such a distinction between $\beta$-sheets and coils, we proceed by
first calculating the partition functions of $\beta$-sheets and coils separately.

Under the action of an external force $f$, the energy of a $\beta$-sheet of
$n_{\beta}$ columns is
\begin{equation}
  \label{eq:beta_energy}
  E_{\beta}=-\epsilon \sum\limits_{j=1}^{n_{\beta}-1} {\rm v}(l_j,l_{j+1}) - n_{\beta} f a_0 + 2 (n_{\beta}-1) \Delta \ ,
\end{equation}
where  $l_j$ is the number of monomers in the $j$-th
column of the $\beta$-sheet, and ${\rm v}(l_j,l_{j+1})=\min(l_j,l_{j+1})-1$. 
The partition function of a $\beta$-sheet  with ${n}{\geq}{4}$ monomers is
\begin{equation}
  \label{eq:beta_Z}
  Z_\beta(n)=\sum\limits_{n_{\beta}=2}^{[n/2]}\sum\limits_{l_j\geq 2} \delta_{l_1+\cdots+l_{n_\beta}}^n p^m s^{2(m-1)} 
  \prod\limits_{j=1}^{m-1} a^{{\rm v}(l_j,l_{j+1})} \ ,
\end{equation}
where $p=e^{f a_0/T}$, $s=e^{-\Delta/T}$, $a=e^{\epsilon/T}$, and $T$ is the temperature.
It is easier to calculate the 
generating function $G_\beta(\zeta)$ of the partition function $Z_\beta (n)$ 
than to calculate $Z_\beta(n)$ directly. After some simple matrix operations \cite{Zhou-etal-2006-appendix01}
we find that
\begin{eqnarray}
  G_\beta(\zeta) & \equiv &  \sum\limits_{n=4}^{\infty} 
  \bigl(\zeta / a \bigr)^{n} Z_{\beta}(n) 
  \label{eq:beta_GZ} \\
  &=&{p^2 s^2 \over a}  \sum\limits_{i,j,k=1} {x_{i} x_{j} A_{i k}
    (\zeta) A_{j k}(\zeta) \lambda_k(\zeta) 
    \over 1 - (p s^2 /a ) \lambda_k(\zeta)} \ .
  \label{eq:beta_GZ_value}
\end{eqnarray}
In Eq.~(\ref{eq:beta_GZ_value}), $x_j=(\zeta/a)^{(j+1)/2}$; 
$\lambda_{1}(\zeta) \geq \lambda_{2}(\zeta)\geq \ldots $ are
the eigenvalues of a $L\times L$ real-symmetric matrix $\Lambda(\zeta)$ 
with elements $\Lambda_{ij}(\zeta)=\zeta^{1+(i+j)/2}/a^{|i-j|/2}$ 
($i,j=1,2,\ldots, L$); and the orthogonal matrix $A(\zeta)$ contains
the eigenvectors of matrix $\Lambda(\zeta)$. The parameter $L$ should be infinity. 
When $\zeta>1$, $\lambda_1(\zeta)=+\infty$ and consequently $G_\beta(\zeta)$ is 
not properly defined.
When $\zeta \leq 1$, all the eigenvalues of matrix $\Lambda(\zeta)$ are finite, 
and the value of 
$G_\beta(\zeta)$ can be calculated by Eq.~(\ref{eq:beta_GZ_value}).  
The point $\zeta=1$ is therefore a singular point of $G_\beta(\zeta)$, where it 
attains a finite 
value as long as $ \lambda_1(1) < a/(p s^2)$. At this point, the first derivative of 
$G_\beta(\zeta)$ with respect to $\zeta$ is \cite{Zhou-etal-2006-appendix01}
\begin{equation}
  G_\beta^\prime(1) \equiv {{\rm d} G_{\beta}(\zeta) / {\rm d} \zeta}|_{\zeta=1} = 
  \lim\limits_{L\to \infty} \sum\limits_{n=1}^{L}  n y(n) + O(1) \ ,
  \label{eq:Gbetadz}
\end{equation}
where $y(n)=\bigl( \sum_{j,k=1}^L A_{n j} A_{k j} x_k \lambda_j / 
(1- p s^2 \lambda_j / a ) \bigr)^2$. 

The configurational energy of a coil segment  is
\begin{equation}
  \label{eq:Ecoil}
  E_c=-n_{\rm c} f a_0 + m_{\rm c} \Delta \ ,
\end{equation}
where $n_{\rm c}$ and $m_{\rm c}$ are, respectively, 
the total number of columns and the total number of bends in the configuration. 
In a coil segment the value of $m_{\rm c}$ strongly depdends on
the configurations. To calculate the partition function $Z_{\rm c}(n)$ of a coil
segment of $n$ monomers, one needs to distinguish among four 
different boundary 
conditions \cite{Zhou-etal-2006-appendix01}. 
The generating function $G_{\rm c}(\zeta)$
of $Z_{\rm c}(n)$ in this case is expressed as
\begin{eqnarray}
& &  G_{\rm c}(\zeta) \equiv  \sum\limits_{n=0}^\infty \bigl( \zeta/a \bigr)^{n} Z_{\rm c}(n) 
  \label{eq:coil_GZ} \\
  & & =  { a s^2 ( a- \zeta ) ( a + \zeta p ) \over a^3 - \zeta a^2 (1+ p) + \zeta^2 a p (1-s^2) - \zeta^3 p^2 s^2} \ . 
  \label{eq:coil_GZ_value}
\end{eqnarray}
The divergence radius of $G_{\rm c}(\zeta)$ is easy to obtain from Eq.~(\ref{eq:coil_GZ_value}). 
$G_{\rm c}(\zeta)$ approaches $+\infty$ as $\zeta$ approaches this divergence radius.

{\em The collapse transition}.---Since every configuration of 
the polymer is of the form $\ldots$-$\beta$-c-$\beta$-c$\ldots$, 
the generating function $G(\zeta)$ of the whole polymer's partition function $Z(N)$ 
can readily be written down \cite{Lifson-1964,Zhou-etal-2006-appendix01}:
\begin{equation}
  G(\zeta)\equiv \sum\limits_{N=0}^\infty \bigl(\zeta / a\bigr)^{N} Z(N) 
  ={{[1+G_{\beta}(\zeta)]G_{\rm c}(\zeta)} \over {1-G_{\beta}(\zeta) G_{\rm c}(\zeta)}} \ .
  \label{eq:GZ_total}
\end{equation}
$Z(N)$ is related to the free energy density $g(f,T)$ of the polymer 
by $Z(N)=\exp\bigl(-N \beta g(f,T)\bigr)$. Therefore, 
in the thermodynamic limit $N\to \infty$, the free energy density is
\begin{equation}
  \label{eq:free_energy_expression}
  g(f,T)= -\epsilon + T \ln \zeta_0  \ ,
\end{equation}
where $\zeta_0$ is the smallest positive root in the 
range $0< \zeta_0 < 1$ of the following equation 
\begin{equation}
  \label{eq:zeta_0}
  G_{\beta}(\zeta_0) G_{\rm c}(\zeta_0)=1 \ .
\end{equation}
If Eq.~(\ref{eq:zeta_0}) has no root in the range of $0 < \zeta_0 <1$, then $\zeta_0=1$. The
polymer's relative extension $\ell$ along ${\bf z}_0$ is
\begin{equation}
  \label{eq:extension}
  \ell \equiv - {\partial \ln \zeta_{0} \over \partial \ln p} = - {1 \over  \beta a_0 \zeta_0 } \  { \partial \zeta_0 \over \partial f} \ .
\end{equation}
A similar expression can also be obtained for the density of contacts. 
For the simplicity we take the relative extension
$\ell$ as our order parameter.

When both temperature $T$ and external force $f$ 
are sufficiently low, Eq.~(\ref{eq:zeta_0}) has
no root in $0 < \zeta_0 < 1$. Then $g(f,T)=-\epsilon$. The polymer is in the $\beta$-sheet phase, and
$\ell \equiv 0$. There is no significant entropic contribution to the free energy
density. As the temperature or the external force is elevated to certain point 
such that Eq.~(\ref{eq:zeta_0}) is satisfied exactly at $\zeta_0=1$,
a phase transition  occurs. 
At this point, the $\beta$-sheet phase change to the extended coil phase. 
According to the work of Lifson \cite{Lifson-1964},
this globule-coil transition is first-order 
if the value of $G_\beta^\prime(1)$ in Eq.~(\ref{eq:Gbetadz}) 
is {\em finite} at the phase transition point; otherwise,
it is a continuous phase transition.  At the collapse phase transition point,
high-precision numerical calculations  reveal that
$y(n)$ of Eq.~(\ref{eq:Gbetadz}) 
decays exponencially with $n$ as long as $\Delta >0$ \cite{Zhou-etal-2006-appendix01}.
In the vicinity of ${\Delta}{=}{0}$ we have
\begin{equation}
  \label{eq:ynscale}
  y(n) \sim e^{- 0.88 n \Delta / \epsilon  } \hspace*{1.0cm} ( \Delta \ll  \epsilon ) \  .
\end{equation}
Therefore, at the phase transition point $G_\beta^\prime(1)$ is finite as long as
$\Delta >0$. In other words, the collapse phase transition is
first-order for polymers with bending stiffness $\Delta >0$;  it  is second-order
for polymers with exactly zero bending stiffness. It is remarkable 
that the cooperativity
of the collapse transition can be changed by adding just a
small bending stiffness.

%%% Results: the case of zero bending stiffness %%%

\begin{figure}[bth]
  \includegraphics[width=0.6\linewidth,angle=270]{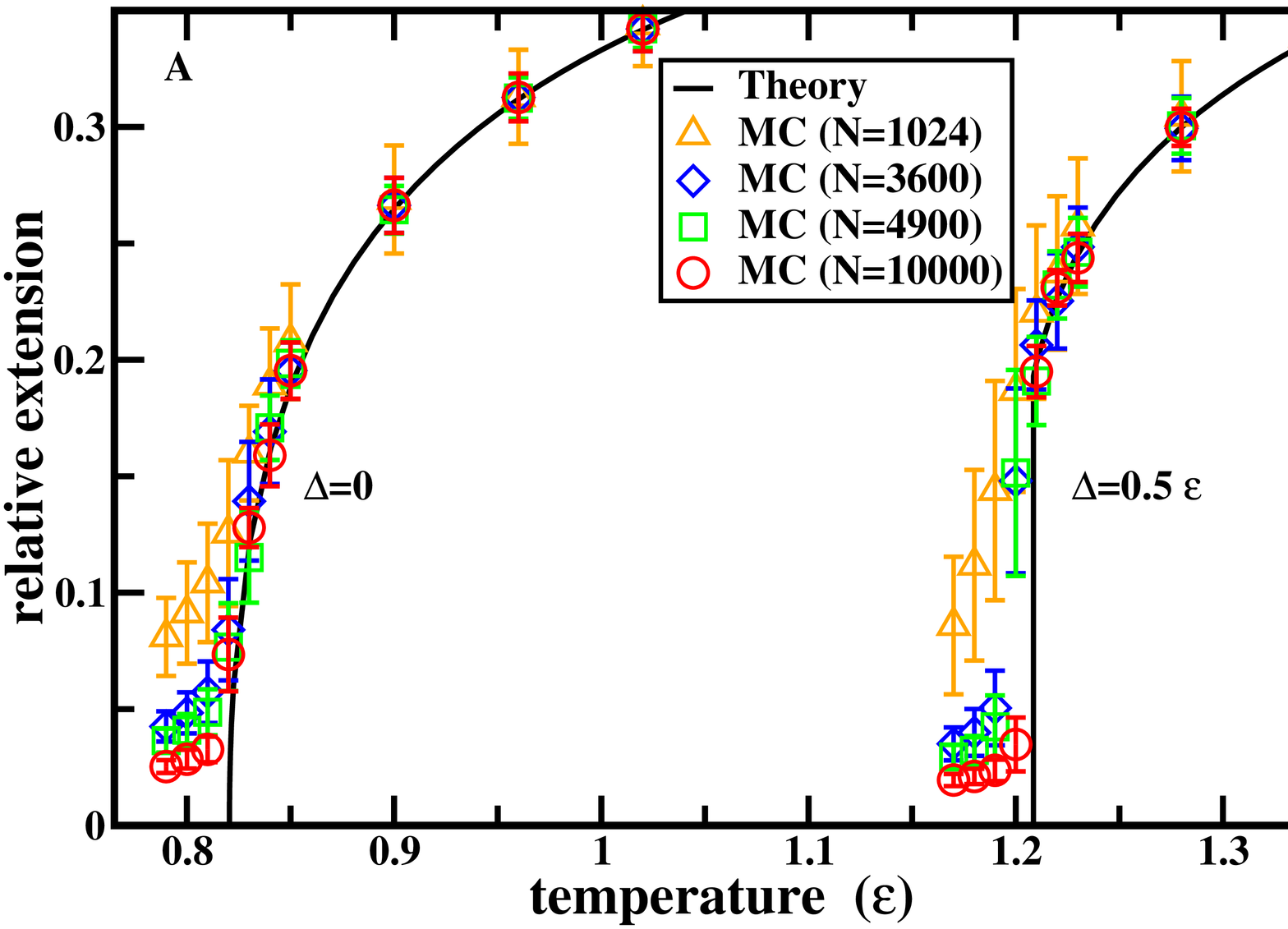}
  \includegraphics[width=0.6\linewidth,angle=270]{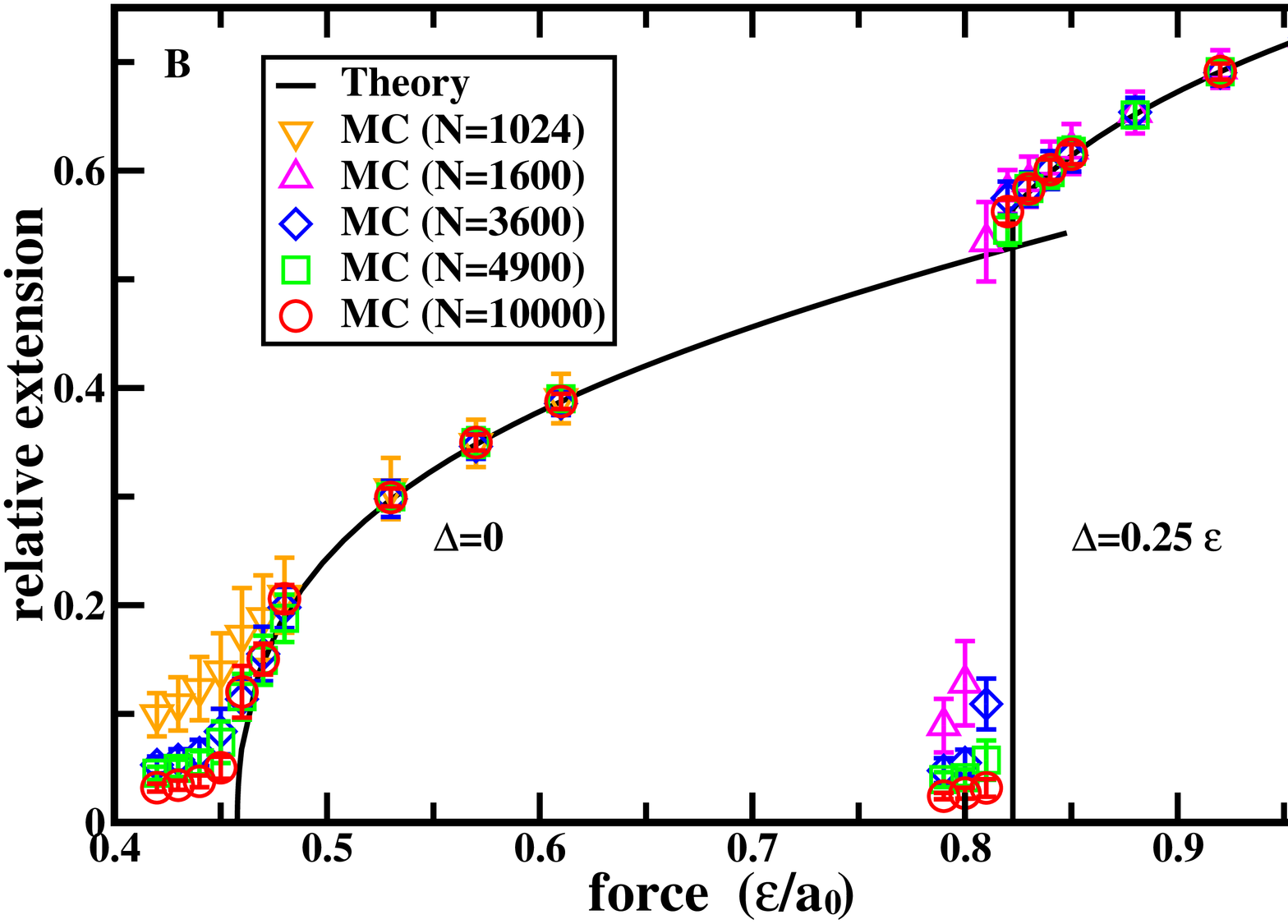}
  \caption{\label{fig:ET_and_EF} 
   (Color online) Response behavior of the polymer. 
   (A) temperature--extension curve
    at force $f=0$;
    (B) Force--extension curve at temperature $T=0.590928 \epsilon$. 
    Solid lines are analytical predictions and  symbols are MC 
    simulation results 
    for different system sizes.
  }
\end{figure}

\begin{figure}[t]
  \includegraphics[width=0.5\linewidth,angle=270]{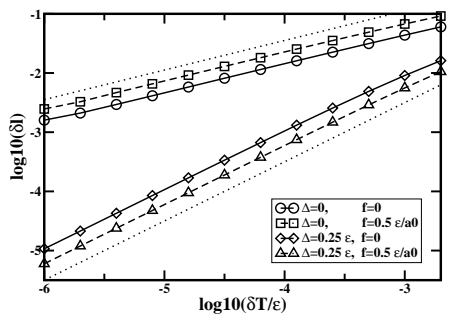}
  \caption{\label{fig:scaling}
    The scaling relation between extension increment $\delta \ell$  
    $[\equiv \ell\bigl( T_{\rm gc}(f)+ \delta T \bigr) - \ell \bigl(T_{\rm gc}(f)\bigr) ]$ 
    and  $\delta T$. The upper and lower dotted line has slope $1/2$ and $1$, respectively.
    The dimension of the matrix $\Lambda(\zeta)$ is set to $L=5000$.
  }
\end{figure}

{\em Results for $\Delta \equiv 0$.} We perform  analytical calculations
as well as Monte Carlo simulations on the model system.
The $\ell$-$T$ relation at $f=0$
is shown in Fig.~\ref{fig:ET_and_EF}a, and the $\ell$-$f$ relation at
$T=0.591 \epsilon$ is shown in Fig.~\ref{fig:ET_and_EF}b. 
When there is no external force, a second-order globule-coil phase transition 
occurs at $T_{\rm gc}(0)=0.8205 \epsilon $. 
The force-induced globule-coil
transition at constant temperature is also second-order. 
Our findings are in accordance 
with Refs.~\cite{Grassberger-Hsu-2002,Rosa-etal-2003}.
In the vicinity of the collapse  transition, the relative
extension of the extended coil phase can be expressed in a scaling form
in terms of temperature change $\delta T$  or force change $\delta f$:
\begin{eqnarray}
  \ell\bigl(T_{\rm gc}(f)+ \delta T\bigr) &\approx& \ell\bigl(T_{\rm gc}(f)\bigr) + c_1 \ (\delta T)^{\gamma_1} \ ,
  \label{eq:ext_scale_T} \\
  \ell\bigl(f_{\rm gc}(T)+ \delta f\bigr) &\approx& \ell\bigl(f_{\rm gc}(T)\bigr) + c_2 \ (\delta f)^{\gamma_2} \ ,
  \label{eq:ext_scale_F}
\end{eqnarray}
where $T_{\rm gc}(f)$ and $f_{\rm gc}(T)$ are, respectively, the transition temperature at 
fixed force and the transition force at fixed temperature;
$\gamma_1$ and $\gamma_2$ are two scaling exponents; and $c_1, c_2$ are two numerical constants. 
In the case of $\Delta \equiv 0$, we have $\ell\bigl(T_{\rm gc}(f)\bigr) =
\ell\bigl( f_{\rm gc}(T)\bigr)=0$; and we find that the scaling exponent in
Eq.~(\ref{eq:ext_scale_T}) is independent of force with $\gamma_1=1/2$ (see Fig.~\ref{fig:scaling}),
which is consistent with a second-order phase transition.
The scaling relation Eq.~(\ref{eq:ext_scale_F}) with $\gamma_2=1/2$ is also confirmed by
numerical calculation (data not shown).

%%% Results: the case of non-zero bending stiffness %%%

{\em Results for $\Delta >0$}. In this case, the collapse transition becomes first-order. 
For example at $\Delta = 0.5 \epsilon $ and $f=0$, the relative extension jumps from zero to
$\ell=0.193$ at the transition temperature $T_{\rm gc}(0)=1.209 \epsilon$ 
(Fig.~\ref{fig:ET_and_EF}a).  Such a large jump is also observed in the
force-induced transition (Fig.~\ref{fig:ET_and_EF}b).
A non-zero bending stiffness therefore is able to dramatically enhance the 
cooperativity of
the globule-coil phase transition. This may be partially understood in the following way.
A positive bending energy significantly decreases the configurational
entropy of a coil segment.  Consequently the globule-coil transition
will occur at higher temperature and higher force, and once the polymer is unfolded
it favors those highly elongated configurations which have fewer bends. 

As is expected for a first-order phase transition,
in the case of $\Delta >0$,  the scaling exponents in the scaling
forms (\ref{eq:ext_scale_T}) and (\ref{eq:ext_scale_F}) are $\gamma_1=\gamma_2=1$ 
(Fig.~\ref{fig:scaling}). 

%%% Conclusion and discussion %%%

{\em Conclusion}.---In this work we studied the collapse transition of a
2D partially directed lattice polymer model using the Lifson approach 
\cite{Lifson-1964}. We devided each
 configuration of the polymer into $\beta$-sheet segments and coil segments and then
calculated their partition functions separately. This approach enables us to 
precisely calculate
the free energy density of the system. 
Previous tricritical field theory \cite{Brak-etal-1992,Owczarek-Prellberg-1994} 
and numerical scaling analysis \cite{Rosa-etal-2003} predicted that the collapse
transition  of a flexible (${\Delta}{\equiv}{0}$) partially directed lattice polymer is 
second-order. We have
confirmed this prediction. Our calculations 
further revealed that, the susceptibility $\chi={\partial \ell / \partial f}$ of the polymer diverges as 
$\chi \propto \bigl(f-f_{\rm gc}(T) \bigr)^{-1/2}$ or 
$\chi \propto \bigl(T-T_{\rm gc}(f) \bigr)^{-1/2}$ as $f$ or $T$ approaches
the phase transition value from above. 
The most important result of the present paper is that,
a non-zero bending stiffness of the polymer changes the 
nature of the collapse transition 
from being second-order to being first-order. 
This conclusion is in agreement
with an earlier exact enumeration study \cite{Kumar-Giri-2005}. 
It is well known that, in the 2D Ising model the paramagnet--ferromagnet 
phase transition changes from being second-order to being first-order under
the action of a non-zero external magnetic field. 
It it is interesting that in the 2D polymer system such a qualitative
change is caused not by an external field (such as the external force), but by an 
internal (microscopic) parameter, the bending stiffness $\Delta$. 

At this point, it remains to be seen whether the above conclusion also holds if we remove
the constraint of partial directedness. In the case of DNA thermal denaturation, including 
long-range excluded-volume interactions between different
bubbles changes the order of the phase transition
\cite{Kafri-etal-2000}.  When additional long-range excluded-volume interactions are
included, it may not be surprising if the temperature-induced 2D collapse
transition will be first-order only when 
the bending energy is much larger than the contacting energy $\epsilon$
\cite{Doniach-etal-1996}. On the other hand, the force-induced 2D collapse transition 
may have different behavior.
Mean-field work of Ref.~\cite{Grassberger-Hsu-2002} suggested that in the globule-coil
transition, when a force is approached from below, a collapsed 2D polymer takes shape of an
elongated elipse (i.e., partially directed). 
The mathematical constraint of partial directedness of the present paper
may not qualitatively change the cooperativity of the force-induced 2D collapse transition.
Additional 
MC simulation work as well as exact numerical enumeration study are
needed to clarify this point. (When $\Delta=0.3 \epsilon$,
our prelimilary MC results hinted at 
a first-order phase transition, with $f_c \approx 0.9 \epsilon$ at $T=0.6 \epsilon$.) 
It is also highly desirable to perform real 2D polymer collapse experiments. 
For example, a flexible or semiflexible
biopolymer [such as DNA and poly(Glycine)] can be attached to a mobile lipid bilayer 
\cite{Maier-Raedler-1999}. Polymer configurations can be recorded in real-time  and 
manipulated by controlling temperature or external force.

%%% Acknowledgement %%%

We thank X.-S. Chen, D.~Giri, P. Grassberger, D.~Marenduzzo, H. Orland for critical
comments, and the State Key Lab.~of Sci.~and Engn.~Computing of CAS for 
computational facilities. 
HZ acknowledges the initial  support of R.~Lipowsky and the 
AvH foundation, SK acknowledges the hospitality of
the MPIPKS-Dresden.

\end{document}